\documentclass[twocolumn,preprintnumbers,amsmath,amssymb]{revtex4}

\ifx\pdfoutput\undefined
\usepackage[dvips]{graphicx}
\else
\usepackage[pdftex]{graphicx}
\fi
\usepackage{dcolumn}
\usepackage{bm}

\begin{document}
\title{Long Delay Times in Reaction Rates Increases Intrinsic Fluctuations}

\author{Matthew Scott}\email{mscott@math.uwaterloo.ca}
\affiliation{Department of Applied Mathematics, University of Waterloo, \\
Waterloo, Ontario, Canada N2L 3G1}

\begin{abstract}
In spatially distributed cellular systems, it is often convenient to represent complicated auxiliary pathways and 
spatial transport by time-delayed reaction rates. Furthermore, many of the reactants appear in low numbers 
necessitating a probabilistic description. The coupling of delayed rates with stochastic dynamics leads to a 
probability conservation equation characterizing a non-Markovian process. A systematic approximation is derived that 
incorporates the effect of delayed rates on the characterization of molecular noise, valid in the limit of long delay 
time. By way of a simple example, we show that delayed reaction dynamics can only \emph{increase} intrinsic 
fluctuations about the steady-state. The method is general enough to accommodate nonlinear transition rates, allowing 
characterization of fluctuations around a delay-induced limit cycle. 
\end{abstract}


\maketitle

\section{Introduction}
Biochemical circuits underlying many complicated cell functions, disease states, or viral propagation are often 
modeled by systems of delayed differential equations~\cite{HIVDelay,MackeyRBC,ZebraFish,SneppenHes1,Kuske2,Barrio}, 
where the delay time represents auxiliary reaction pathways or spatial transport. It is well-known that in chemical 
reaction networks at the cellular level, many of the reactants are present in low copy number and therefore intrinsic 
noise is simply one of the operating constraints~\cite{Kerszberg}. It is, however, difficult to develop and analyze 
models that contain both stochastic dynamics and delayed reaction rates. 

For systems without delay, a stochastic description typically takes the form of a chemical Master equation that 
governs the probability distribution $P({\bf n},t)$ of finding the system in state ${\bf n}$ at time $t$ conditioned 
upon some initial state~\cite{ChemMaster,Gillespie1992}. The Master equation can rarely be solved exactly, and 
various methods have been developed to approximate the evolution of $P({\bf n},t)$. Perhaps the most well-known is 
the Kramers-Moyal expansion, which when truncated after the second term results in a diffusion equation with 
nonlinear drift and non-constant diffusion, called the Fokker-Planck equation~\cite{Gillespie2,GardinerBOOK}. If the 
deterministic system evolves along a stable trajectory near a stable fixed point, van Kampen~\cite{VanKampenME} has 
developed an alternate approximation of the Master equation that relies upon a perturbation expansion in some 
extensive quantity, providing a consistent characterization of the fluctuations in terms of a Fokker-Planck equation 
with linear drift and constant diffusion from whence the mean and variance are easily computed. (For a more detailed 
discussion, see~\cite{VanKampenME,Gillespie2} and references therein.) 

Nevertheless, in many systems the individual reaction events depend upon the past state of the 
system~\cite{Kerszberg,Bratsun}, and the methods developed to approximate the solution of the Master equation are no 
longer appropriate. Writing the transition probability of moving from state ${\bf n}'$ to state ${\bf n}$ in an 
interval $dt$ as $W_{{\bf n}',{\bf n}} dt$ and the two-point joint probability distribution of finding the system in 
state ${\bf n}$ at time $t$ and in state ${\bf m}$ at time \mbox{$t-\tau$} as \mbox{$P_2({\bf n},t;{\bf m},t-\tau)$}, 
the delayed dynamics introduce a convolution term into the probability conservation equation,
\begin{gather}
\label{eq:ConvMasterEq}
\frac{\partial P({\bf n},t)}{\partial t}=\sum\limits_{{\bf n}'}{W_{{\bf n}',{\bf n}}P({\bf n}',t)-W_{{\bf n},{\bf 
n}'}P({\bf n},t)}+\\
\sum\limits_{{\bf n}'}{\sum\limits_{{\bf m}}{W^{\tau}_{{\bf n}',{\bf n}}P_2({\bf n}',t;{\bf m},t-\tau)-W^{\tau}_{{\bf 
n},{\bf n}'}P_2({\bf n},t;{\bf m},t-\tau)}}.\notag
\end{gather}
Here $W^{\tau}_{{\bf n}',{\bf n}}({\bf m})$ depends upon the state at a time $\tau$ in the past: ${\bf m}={\bf 
n}(t-\tau)$. Eq.~\ref{eq:ConvMasterEq} is no longer a closed equation for $P({\bf n},t)$ since it now includes the 
unknown distribution \mbox{$P_2({\bf n},t;{\bf m},t-\tau)$}. As a consequence, it no longer describes a Markov 
process and methods used to treat the standard chemical Master equation require modification.

Many investigations using stochastic simulation algorithms have illustrated the importance of intrinsic noise in 
systems with delay~\cite{ZebraFish,Kerszberg,Barrio}, while past analytic work has focused primarily upon 
approximations of the delayed nonlinear Fokker-Planck equation~\cite{LongtinDelay,Frank}, stochastic delay 
differential equations~\cite{MackeyDelay,Kuske,Kuske2} or exactly-solvable random-walk 
models~\cite{MiltonDelay,MiltonBOOK}. While each provides considerable insight into the interdependent effects of 
noise and time delay, comparatively little work has been done to connect the underlying discrete probability 
conservation equation to these continuous approximations. In what follows a perturbation scheme is developed that, 
under the condition that the delay time exceeds the relaxation time of the deterministic system, allows a general 
probability conservation equation to be approximated by a delayed linear Fokker-Planck equation, thereby making 
connection to past studies. Recent work by Bratsun {\it et al.}~\cite{Bratsun} has explored very similar questions, 
although their method is applicable only to systems with linear reaction rates. The delayed linear noise 
approximation, which is an extension of van Kampen's linear noise approximation~\cite{VanKampenME}, provides a 
consistent characterization of intrinsic fluctuations in delayed systems and is sufficient to show that under fairly 
general conditions delayed reaction events can only increase the magnitude of these fluctuations. A simple example of 
a birth/death process is used to provide a concrete implementation of the method, and a nonlinear predator-prey model 
illustrates characterization of fluctuations along a delay-induced limit cycle.

\section{Mathematical methods}
\label{sec:methods}

A stochastic model of a network of chemical reactions governs the probability distribution $P({\bf n},t)$ of finding 
the system in state ${\bf n}$ at time $t$, with dynamics given in terms of the stoichiometric change resulting from 
the completion of each reaction and the propensity of occurrence for each reaction event, recorded, respectively, in 
the stoichiometry matrix ${\bf S}$ and the propensity vector $\tilde{\bm{\nu}}$~\cite{Gillespie,Elf}. Consider a 
system with $N$ reactants that can combine through one of $R$ reactions. To facilitate the inclusion of delayed 
kinetics into the formalism, we separate the $R$ reactions into two groups: those with rates $\left\{ {\tilde \nu _j 
\left( {\mathbf{n}} \right)} \right\}_{j = 1}^{R_c } $ depending upon the current state of the system and those with 
rates $ \left\{ \varepsilon{\tilde \nu _j \left( {{\mathbf{m}},{\mathbf{n}}} \right)} \right\}_{j = R_c  + 1}^R $  
depending upon the past state of the system $ {\mathbf{m}} = {\mathbf{n}}\left( {t - \tau } \right)$, where $\tau   $ 
is the delay time. The parameter $\varepsilon$ measures the delayed feedback strength; throughout, we shall assume 
that the feedback is weak, and explicitly retain leading-order terms in $\varepsilon$. 

To keep the notation compact, we introduce the step-operator ${\bf E}_i^{-S_{ij}}$ that acts to increment the 
variable $n_i$ by an integer $-S_{ij}$: ${\bf E}_i^{-S_{ij}}f(n_i)= f(n_i-S_{ij})$. Denoting the system volume by 
$V$, Eq.~\ref{eq:ConvMasterEq} takes the form \cite{vanKampenBOOK,Elf,Bratsun},
\begin{widetext}
\begin{equation}
\frac{{\partial P}}
{{\partial t}} = V\mathop {\sum\limits_{j = 1}^{R_c } {\left[ {\left( {\prod\limits_{i = 1}^N {{\mathbf{E}}_i^{ - 
S_{ij} } } } \right) - 1} \right]\tilde \nu _j \left( {\mathbf{n}} \right)P\left( {{\mathbf{n}},t} \right)} 
}\limits_{{\text{current - state dynamics}}}  + V\mathop {\sum\limits_{j = R_c  + 1}^R {\left[ {\left( 
{\prod\limits_{i = 1}^N {{\mathbf{E}}_i^{ - S_{ij} } } } \right) - 1} \right]\sum\limits_{\mathbf{m}} {\varepsilon 
\tilde \nu _j \left( {{\mathbf{m}},{\mathbf{n}}} \right)\Theta \left( {\left\{ {n_i } \right\}} \right)P_2 \left( 
{{\mathbf{n}},t;{\mathbf{m}},t - \tau } \right)} } }\limits_{{\text{delayed - state dynamics}}} ,
\label{eq:MasterEq}
\end{equation}
\end{widetext}
where $\Theta \left( {\left\{ {n_i } \right\}} \right)$ is the Heaviside-step function that ensures no delayed 
reaction occurs if completion results in the unphysical end-state $n_i<0$ for any elements of ${\bf n}$. Throughout, 
only \emph{non-consuming} reactions are considered~\cite{Barrio,CaiDSSA}, {\it i.e.} reactants of an unfinished 
reaction are allowed to participate in new reactions.

The solution of the full distribution $ P\left( {{\mathbf{n}},t} \right)$ is not possible in general, therefore we 
seek an approximate solution. To that end, we make the ansatz that the number of reactant molecules is large enough 
that the discrete molecule numbers $n_i$ can be represented by the continuous deterministic concentrations $x_i$ and 
some continuous fluctuations $\alpha_i$  that scale as the square-root of the number of molecules~\cite{VanKampenME},
\begin{gather}
n_i  = V x_i  + \sqrt V  \alpha _i \quad \mbox{and} \quad m_i  = V x_i^{\tau   }  + \sqrt V  \beta _i ,
\label{eq:change}
\end{gather}
where $V$ is the system volume and $x_i^{\tau   }  \equiv x_i \left( {t - \tau   } \right)$. Using the auxiliary 
variable $\beta_i$, the delayed fluctuations are written as $\beta _i \left( {t - \tau   } \right) = \alpha _i \left( 
{t - \tau   } \right)$ to emphasize the approximation made below; specifically, that the delay time is sufficiently 
large that $\alpha_i$ and $\beta_i$ can be treated as independent random functions. As we show below 
(Eq.~\ref{eq:AutoCorrExpl}), that assumption is consistent with the requirement that the delay time is much larger 
than the characteristic relaxation time of the deterministic equations and that the delayed feedback is weak 
($\varepsilon \ll 1$). In the expansion below, $1/V$ is assumed small, although since $x_i$ is held fixed, 
(equivalently, one assumes that $n_i$ is large). The resulting approximation will be called the \emph{delayed linear 
noise approximation}.

Invoking the linear noise approximation by Taylor-expanding the microscopic transition rates about the macroscopic 
trajectory ${\mathbf{x}}\left( t \right)$ in powers of $\sqrt V$~\cite{VanKampenME,Elf}, we have
\begin{gather*}
\tilde \nu _j \left( {{\mathbf{m}},{\mathbf{n}},V } \right) \approx \nu _j \left( {{\mathbf{x}}^{\tau   } 
,{\mathbf{x}}} \right) + \frac{1}{\sqrt V}  \sum\limits_{i = 1}^N {\left[ {\frac{{\partial \nu _j }}
{{\partial x_i }}\alpha _i  + \frac{{\partial \nu _j }}
{{\partial x_i^{\tau   } }}\beta _i } \right]},
\end{gather*}
with an analogous expression for $\tilde \nu _j \left( {{\mathbf{n}},V } \right)$. The rates $\nu_j$ correspond to 
the deterministic reaction rates (see \cite{Elf} for a discussion of the difference between $\tilde{\bm \nu}$ and 
${\bm \nu}$). The discrete step-operator $\mathbf{E}$ is likewise expressed as a Taylor series in 
$\sqrt{V}^{-1}$~\cite{VanKampenME,Elf},
\begin{gather*}
\prod\limits_{i = 1}^N {{\mathbf{E}}_i^{ - S_{ij} } }  = 1 - \frac{1}
{{\sqrt V  }}\sum\limits_{i = 1}^N {S_{ij} \partial _i }  + \frac{1}
{{2V }}\sum\limits_{i,k = 1}^N {S_{ij} S_{kj} \partial _{i}\partial_{k} } ,
\end{gather*}
where $\partial _i  = {\partial  \mathord{\left/{\vphantom {\partial  {\partial \alpha _i }}} 
\right.\kern-\nulldelimiterspace} {\partial \alpha _i }}$. 
The one-point $P\left( {{\mathbf{n}},t} \right)$ and two-point joint probability $P_2 \left( 
{{\mathbf{n}},t;{\mathbf{m}},t - \tau   } \right)$ can be written in terms of the single distribution and joint 
distribution of the fluctuations about the macroscopic trajectory, $\Pi\left( {\bm{\alpha} ,t } \right)$ and $
\Pi _2 \left( {\bm{\alpha} ,t;\bm{\beta} ,t - \tau   } \right)$, respectively, via the linear change of variables 
suggested by Eq.~\ref{eq:change},
\begin{gather}
P\left( {{\mathbf{n}},t} \right) = V ^{ - N/2} \Pi\left( {\bm{\alpha} ,t } \right),\\
P_2 \left( {{\mathbf{n}},t;{\mathbf{m}},t - \tau   } \right) = V ^{ - N} \Pi _2 \left( {\bm{\alpha} ,t;\bm{\beta} ,t 
- \tau   } \right),
\end{gather}
where ${\bm \alpha}$ and ${\bm \beta}$ are centered upon ${\mathbf{x}}\left( t \right)$ and ${\mathbf{x}}\left( {t - 
\tau   } \right)$, respectively, and the factor involving $V$ comes from the normalization of $\Pi$,
\begin{gather}
\int_{-\infty}^{\infty}P\left( {{\mathbf{n}},t} \right)d{\mathbf{n}}=V ^{ - N/2}\int_{-\infty}^{\infty}\Pi\left( 
{\bm{\alpha} ,t } \right)d{\bm{\alpha}}=1.
\label{eq:NormalizationPi}
 \end{gather}
The decoupling of the deterministic component ${\bm x}$ from the stochastic component ${\bm \alpha}$, with a 
$1/\sqrt{V}$ scaling of the fluctuations implied by the ansatz Eq.~\ref{eq:change}, is the most fundamental step in 
the approximation. That assumption leads directly to the normalization above, and allows subsequent terms in the 
perturbation expansion to be ordered in terms of powers of $1/\sqrt{V}$~\cite{VanKampenME}.

For long delay time $\tau$, and small delayed-feedback strength $\varepsilon$, the fluctuations at time $t$ are 
approximately independent of the fluctuations at time $t - \tau   $, allowing the joint-distribution to be factored 
as,
\begin{gather}
\label{eq:separate}
\Pi _2 \left( {\bm{\alpha} ,t;\bm{\beta} ,T   } \right) \approx \Pi \left( {\bm{\alpha} ,t} \right) \times \Pi \left( 
{\bm{\beta} ,T   } \right)+\varepsilon h({\bm \alpha}, {\bm \beta})\\\notag \quad T \le t-\tau, 
\end{gather}
where $h({\bm \alpha},{\bm \beta})$ must obey the consistency condition,
\begin{gather*}
\int h({\bm \alpha},{\bm \beta})d{\bm \alpha}=\int h({\bm \alpha},{\bm \beta})d{\bm \beta}=0.
\end{gather*}
Notice the factoring of the fluctuations $\Pi_2$ is \emph{not} equivalent with assuming independence in the full 
state, $P({\bf n},t;{\bf m},\tau)=P({\bf n},t)\times P({\bf m},\tau)$ (as is done in~\cite{Bratsun}). Independence of 
the full state is inconsistent with the deterministic rate equations,
\begin{gather}
\frac{{d{\mathbf{x}}}}
{{dt}} = {\mathbf{S}} \cdot \bm{\nu} ={\mathbf{f}}\left( {{\mathbf{x}}^{\tau   } ,{\mathbf{x}}} \right),
\label{eq:detEqs}
\end{gather}
where, in the limit of large ${\bf n}$, the present state is \emph{completely} determined by the past-states (except, 
perhaps, at steady-state). Moreover, independence of the full state implies that higher-order correlations, including 
the autocorrelation function $K(t-s)$, vanish for $|t-s|>\tau$. We show below (Eq.~\ref{eq:AutoCorrExpl}) that this 
is not the case.

In the limit $1/V\to 0 \;(n_i\to \infty)$, the Heaviside step function is $\Theta \left( {\left\{ {n_i } \right\}} 
\right)=1$, and with the factored joint distribution, Eq.~\ref{eq:separate}, the sum over ${\bf m}$ is replaced by 
the integral over ${\bm \beta}$,
\begin{widetext}
\begin{gather*}
 \int\limits_{ - \infty }^\infty  {\left\{ {\nu _j  + \frac{1}{\sqrt V}  \sum\limits_{i = 1}^N {\left[ 
{\frac{{\partial \nu _j }}
{{\partial x_i }}\alpha _i  + \frac{{\partial \nu _j }}
{{\partial x_i^{\tau   } }}\beta _i } \right]} } \right\}V ^{ - N} \Pi \left( {\bm{\alpha} ,t} \right)\Pi \left( 
{\bm{\beta} ,t - \tau   } \right)d{\bm \beta} } = \left\{ {\nu _j + \frac{1}{\sqrt V}  \sum\limits_{i = 1}^N {\left[ 
{\frac{{\partial \nu _j }}
{{\partial x_i }}\alpha _i  + \frac{{\partial \nu _j }}
{{\partial x_i^{\tau   } }}\left\langle {\beta _i } \right\rangle } \right]} } \right\}V ^{ - \frac{N}
{2}} \Pi \left( {\bm{\alpha} ,t} \right).
\end{gather*}
\end{widetext}
Here, the first two terms on the right-hand side follow from the normalization condition on $\Pi(\bm{\beta},t-\tau)$, 
Eq.~\ref{eq:NormalizationPi}, and the third from,
\begin{gather*}
V^{-N/2}\int_{-\infty}^{\infty}\beta_i\Pi(\bm{\beta},t-\tau)d\bm{\beta}=\langle \beta_i \rangle.
\end{gather*}

Substituting the expanded terms in Eq.~\ref{eq:MasterEq}, using the chain-rule to write the partial derivatives of 
$P({\bf n},t)$ in terms of $\Pi$ and ${\bm \alpha}$~\cite{VanKampenME}, and collecting in powers of $\sqrt V$, the 
zero'th order term is simply the deterministic delayed reaction rate equations, Eq.~\ref{eq:detEqs}. At ${\sqrt 
V}^{-1} $, we obtain the equation characterizing the probability distribution of the fluctuations ${\bm \alpha}(t)$,
\begin{gather}
\begin{split}
\frac{{\partial \Pi }}
{{\partial t}} =  \sum\limits_{i,j} {-\Gamma_{ij} \partial _i \left( {\alpha _j \Pi } \right)  +  \frac{D_{ij}}{2} 
\partial _{ij} \Pi  - \varepsilon\Gamma_{ij}^{\tau   } \left\langle {\beta_j } \right\rangle \partial _i \Pi } ,
\end{split}
\label{eq:delayFP}
\end{gather}
where,
\begin{gather*}
\Gamma_{ij}  = \frac{{\partial \lbrack {\bm S}\cdot {\bm \nu}\rbrack_i }}
{{\partial x_j }}, \;
\varepsilon\Gamma_{ij}^{\tau   }  = \frac{{\partial \lbrack {\bm S}\cdot {\bm \nu}\rbrack_i }}
{{\partial x_j^{\tau   } }}, \; {\bm D} = {\mathbf{S}} \cdot \mbox{diag}\left[ \bm{\nu}  \right] \cdot 
{\mathbf{S}}^T.
\end{gather*}
Eq.~\ref{eq:delayFP} is a closed diffusion equation for $\Pi({\bm \alpha},t)$ with coefficients that are linear in 
the fluctuation variables ${\bm \alpha}$. The matrix ${\bm D}$ represents the diffusive effects of the fluctuations, 
while the matrices ${\bm \Gamma}$ and ${\bm \Gamma^{\tau}}$ represent the restorative drift in the 
system~\cite{MeBrianMads,vanKampenBOOK}. Eq.~\ref{eq:delayFP} is not quite a Fokker-Planck equation since it contains 
the delayed average $\langle \beta_j \rangle \equiv \langle \alpha_j (t-\tau) \rangle$ in the drift coefficient. 
Nevertheless, the initial conditions can be chosen so that the last term in Eq.~\ref{eq:delayFP} vanishes, as we now 
show. Multiplying $\frac{{\partial \Pi }}
{{\partial t}}$ by $\alpha_i$ and integrating yields the evolution equation for the mean,
\begin{gather}
\frac{{d\left\langle {\bm{\alpha} \left( t \right)} \right\rangle }}
{{dt}} = {\bm \Gamma} \cdot \left\langle {\bm{\alpha} \left( t \right)} \right\rangle  + \varepsilon{\bm 
\Gamma}^{\tau   }  \cdot \left\langle {\bm{\alpha} \left( {t - \tau   } \right)} \right\rangle .
\label{eq:MeanEq}
\end{gather}
The initial average $\langle \bm{\alpha} (t) \rangle$ can always be absorbed into the initial conditions on 
${\mathbf{x}}\left( t \right)$ so that $\left\langle {\bm{\alpha} \left( t \right)} \right\rangle  = 0$ for $ t 
\leqslant 0$, thereby ensuring that $ \left\langle {\bm{\alpha} \left( t \right)} \right\rangle  = 0$ for all time. 
Without loss of generality, then, we write Eq.~\ref{eq:delayFP} as the Fokker-Planck equation with coefficients 
linear in $\bm{\alpha}$,
\begin{gather}
\frac{{\partial \Pi }}
{{\partial t}} =  - \sum\limits_{i,j} {\Gamma_{ij} \partial _i \left( {\alpha _j \Pi } \right)}  + 
\frac{1}{2}\sum\limits_{i,j} {D_{ij} \partial _{i}\partial_{j} \Pi } .
\label{eq:FokkerPlanck}
\end{gather}
It is important to note that although the fluctuations at time $t$ are independent of fluctuations in the past, they 
are conditioned by the macroscopic solution $ {\mathbf{x}}\left( {t} \right)$ (and ${\mathbf{x}}\left( {t-\tau} 
\right)$) through the coefficient matrices ${\bm \Gamma}$ and ${\bf D}$. 

A consequence of Eq.~\ref{eq:FokkerPlanck} is that, to $O\left( {V ^{ - 1} } \right)$  , the fluctuations are 
Gaussian distributed with covariance $ \Xi_{ij}  = \left\langle {\alpha _i \alpha _j } \right\rangle  - \left\langle 
{\alpha _i } \right\rangle \left\langle {\alpha _j } \right\rangle  = \left\langle {\alpha _i \alpha _j } 
\right\rangle $. Multiplying Eq.~\ref{eq:FokkerPlanck} by $\alpha _i \alpha _j$ and integrating over all ${\bm 
\alpha}$ yields a dynamic equation governing ${\bm \Xi}$~\cite[p. 211]{vanKampenBOOK},
\begin{gather}
\frac{d{\bm \Xi}}{dt}={\bm \Gamma} \cdot {\bm \Xi} + {\bm \Xi} \cdot {\bm \Gamma}^{\dagger} +{\bf D},
\label{eq:LNACorrEv}
\end{gather}
(where ${\bm \Gamma}^\dagger$ is the matrix transpose of ${\bm \Gamma}$; not to be confused with ${\bm 
\Gamma}^\tau$). At steady-state, the coefficient matrices ${\bm \Gamma}$ and ${\bf D}$ (and therefore ${\bm \Xi}$) 
will be constant, satisfying the fluctuation-dissipation relation,
\begin{gather}
{\bm \Gamma} \cdot {\bm \Xi} + {\bm \Xi} \cdot {\bm \Gamma}^\dagger +{\bf D} = 0.
\label{eq:FlucDiss}
\end{gather}
The diffusion matrix ${\bf D}$ is symmetric and positive semi-definite by construction, so that a steady-state 
probability distribution is only possible if the drift term ${\bm \Gamma}$ balances the diffusion ${\bf D}$. With 
long-delay in the reaction kinetics, the restorative influence of ${\bm \Gamma}^\tau$ no longer appears in the 
equation governing the fluctuations (Eq.~\ref{eq:FokkerPlanck}), so that although the delayed dynamics increase the 
magnitude of the diffusion matrix ${\bm D}$, the dissipation due to ${\bm \Gamma}^\tau$ is lost. Therefore, \emph{in 
the limit of long delay time, delayed dynamics can only serve to increase the magnitude of intrinsic fluctuations}. 

\section{Steady-state autocorrelation function and spectrum}

The fluctuation-dissipation relation, Eq.~\ref{eq:FlucDiss}, and the evolution equation for the mean, 
Eq.~\ref{eq:MeanEq}, together provide an expression for the time-autocorrelation matrix for the fluctuations about 
the steady-state, ${\bf K}(t)=\langle {\bm \alpha}(t)\cdot{\bm \alpha}^T(0)\rangle$. The steady-state autocorrelation 
function ${\bf K}(t)$ is, by definition,
\begin{gather*}
{\mathbf{K}}\left( t \right) = \iint {{\bm \alpha} '\left( t \right) \cdot {\bm \alpha} ^T \left( 0 \right)\Pi _2 
\left( {{\bm\alpha} ',t;{\bm \alpha} ,0} \right)d{\bm \alpha} 'd{\bm \alpha} }.
\end{gather*}
Re-writing in terms of the conditional probability,
\begin{gather}
{\mathbf{K}}\left( t \right) = \iint {{\bm \alpha} '\left( t \right) \cdot {\bm \alpha} ^T \left( 0 \right)\Pi \left( 
{{\bm \alpha} ',t|{\bm \alpha} ,0} \right)\Pi \left( {{\bm \alpha} ,0} \right)d{\bm \alpha} 'd{\bm \alpha} }\notag\\
{\mathbf{K}}\left( t \right) = \iint {\left\langle {{\bm \alpha} \left( t \right)} \right\rangle _{{\bm \alpha} 
\left( 0 \right)}  \cdot {\bm \alpha} ^T \left( 0 \right)\Pi \left( {{\bm \alpha} ,0} \right)d{\bm \alpha} },
\label{eq:CondProb}
\end{gather}
where ${\left\langle {{\bm \alpha} \left( t \right)} \right\rangle _{{\bm \alpha} \left( 0 \right)} }$ is the 
solution of Eq.~\ref{eq:MeanEq} with initial condition ${\bm \alpha}(0)$, and $\Pi$ is the equilibrium distribution 
of ${\bm \alpha}(0)$~\cite{Bratsun,NGvKFBook}. Eq.~\ref{eq:separate} requires that the conditional probability 
density $\Pi({\bm \alpha}',t|{\bm \alpha},0)$ be written as a perturbation expansion in $\varepsilon$,
\begin{gather}
\Pi({\bm \alpha}',t|{\bm \alpha},0)=\Pi^0({\bm \alpha}',t|{\bm \alpha},0)+\varepsilon \Pi^1({\bm \alpha}',t|{\bm 
\alpha},0)+O(\varepsilon^2).
\end{gather}
Consequently, in the conditional average ${\left\langle {{\bm \alpha} \left( t \right)} \right\rangle _{{\bm \alpha} 
\left( 0 \right)} }$, only terms to first-order in $\varepsilon$ are retained. 

The conditional average ${\left\langle {{\bm \alpha} \left( t \right)} \right\rangle _{{\bm \alpha} \left( 0 \right)} 
}$ is obtained from Eq.~\ref{eq:MeanEq}, which is easily solved via Laplace transform. The equilibrium correlation 
function is an even function of time-difference alone, equivalent to boundary condition $\left\langle {{\bm \alpha} 
\left( t \right)} \right\rangle=0$ for $t<0$, leading to the formal solution
\begin{gather}
\left\langle {\hat {\bm \alpha} \left( s \right)} \right\rangle_{\alpha \left( 0 \right)}   = \left[ {s{\mathbf{I}} - 
{\bf \Gamma}  - \varepsilon e^{ - s\tau } {\bf \Gamma}^\tau  } \right]^{ - 1}  \cdot \left\langle {{\bm \alpha} 
\left( 0 \right)} \right\rangle .
\label{eq:BratMean}
\end{gather}
The derivation of Eq.~\ref{eq:FokkerPlanck} assumes $\varepsilon \to 0$, so to remain consistent, we retain only 
leading-order terms in $\varepsilon$ in the Laplace transform of the mean,
\begin{gather}
\left\langle {\hat {\bm \alpha} \left( s \right)} \right\rangle  = \left[ {s{\mathbf{I}} - {\bf \Gamma} } \right]^{ - 
1}  \cdot \left\langle {{\bm \alpha} \left( 0 \right)} \right\rangle\notag \\
 + \varepsilon e^{ - s\tau } \left[ {s{\mathbf{I}} - {\bf \Gamma} } \right]^{ - 1}  \cdot {\bf \Gamma}^\tau   \cdot 
\left[ {s{\mathbf{I}} - {\bf \Gamma}} \right]^{ - 1}  \cdot \left\langle {{\bm \alpha} \left( 0 \right)} 
\right\rangle.
\label{eq:myMean}
\end{gather}
With substitution into Eq.~\ref{eq:CondProb}, using the fluctuation-dissipation relation, Eq.~\ref{eq:FlucDiss}, the 
autocorrelation function is
\begin{gather}
\label{eq:AutoCorrExpl}
{\bf K}(t)=\exp\left[{{\bf \Gamma} t}\right]\cdot{\bm \Xi}_s\\
+\varepsilon \Theta(t-\tau)\mathfrak{L}_{t-\tau}^{ - 1} \left\{ {\left[ {s{\mathbf{I}} - {\bf \Gamma} } \right]^{ - 
1}  \cdot {\bf \Gamma}^\tau   \cdot \left[ {s{\mathbf{I}} - {\bf \Gamma} } \right]^{ - 1} } \right\}\cdot{\bm 
\Xi}_s,\notag
\end{gather}
where $\Theta(t-\tau)$ is the Heaviside step function, and $\mathfrak{L}_{t-\tau}^{ - 1}$ is the time-shifted inverse 
Laplace transform. The first term produces an exponential drop from $t=0$, while the second term produces an 
anti-correlated second peak slightly beyond $t=\tau$; higher-order terms in $\varepsilon$ produce alternating 
correlated/anti-correlated peaks of magnitude $O(\varepsilon^n)$ for the $n^{th}$ peak. 

The fluctuation spectrum follows immediately from the autocorrelation function. We denote the 
$\varepsilon$-correction to the autocorrelation function $\hat {\bm K}_{corr}(s)$,
\begin{gather}
\label{eq:corrSpec}
\hat {\bm K}_{corr}(s)=e^{ - s\tau } \left[ {s{\mathbf{I}} - {\bm \Gamma} } \right]^{ - 1}  \cdot {\bm \Gamma}^\tau   
\cdot \left[ {s{\mathbf{I}} - {\bm \Gamma} } \right]^{ - 1},
\end{gather}
then the spectrum $S(\omega)$ is~\cite{GardinerBOOK},
\begin{gather}
\label{eq:spec}
S(\omega)=\int_{0}^{\infty} e^{-i\omega t}{\bf K}(t)dt+\int_{-\infty}^{0} e^{i\omega t}{\bf K}^\dagger(-t)dt\notag\\
=\left[ { - i\omega {\mathbf{I}} + {\bm \Gamma} } \right]^{ - 1}  \cdot {\mathbf{D}} \cdot \left[ {i\omega 
{\mathbf{I}} + {\bm \Gamma}^\dagger } \right]^{ - 1} \notag\\
+\varepsilon \left\{ {{\mathbf{K}}_{corr} \left( {i\omega } \right) \cdot \Xi _s  + \Xi _s  \cdot 
{\mathbf{K}}_{corr}^\dagger \left( { - i\omega } \right)} \right\},
\end{gather}
where ${\bf D}$ is the diffusion matrix introduced in Eq.~\ref{eq:delayFP} evaluated at the steady-state. 

\begin{figure}
\includegraphics[width=\linewidth]{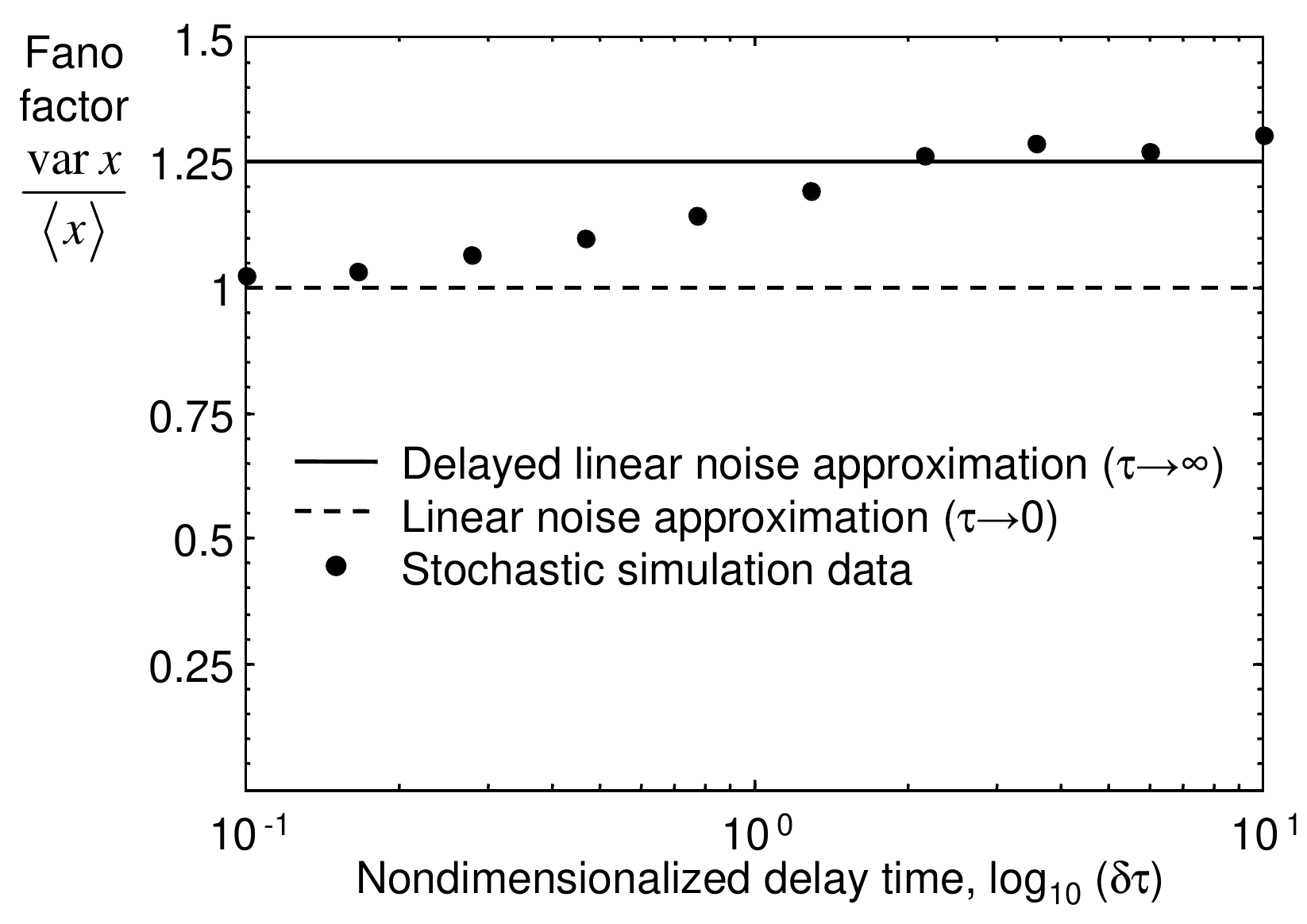}%
\caption{{\bf Fano factor as a function of delay time}. The $\tau=0$ (dotted) and $\tau \to \infty$ (solid) 
approximations to the Fano factor, $\mbox{var}[x]/\langle x\rangle$ (Eq.~\ref{eq:FanoFac}), provide lower and upper 
bounds on the estimate of intrinsic noise over the range of delay times, as compared to stochastic simulation data 
generated from the model shown in Eq.~\ref{eq:tslRates} (filled circles). The delayed linear noise approximation 
adequately characterizes the fluctuations even when the delay time is of the order of the other time scales in the 
problem ({\it i.e.} $\tau \delta \approx 1$). Here, the deterministic reaction rates are $\gamma=100$, $\delta=4$, 
$\varepsilon=1/4$, with $V=1$. Simulation data is from an ensemble of $10^5$ members using the delayed Gillespie 
algorithm~\cite{CaiDSSA}.}
\label{fig:Tau}
\end{figure}

\section{Linear example - Delayed protein degradation}

To provide some context for the formal derivation above, consider a simple birth/death model with delayed 
degradation~\cite{Bratsun}. The total number of species $X$ evolves via the following three reactions,
\begin{gather}
\begin{array}{*{20}c}
   {\mbox{constant synthesis:}} & {X\overset{\nu _1 }{\rightarrow}X + 1;} & {\nu _1  = \gamma},  \\
   {\mbox{linear degradation:}} & {X\overset{\nu _2 }{\rightarrow}X - 1;} & {\nu _2  = \delta\cdot x},  \\
   {\mbox{delayed degradation:}} & {X\overset{\nu _3 }{\Rightarrow}X - 1;} & {\nu _3  = \varepsilon\delta\cdot 
x^{\tau}}.  \\
 \end{array}
 \label{eq:tslRates} 
\end{gather}

The reaction rate vector (in units of concentration/time) is given by \mbox{${\bm \nu}=\lbrack{\gamma,\delta \cdot x, 
\varepsilon\delta \cdot x^{\tau}}\rbrack$} and the stoichiometry matrix is ${\bf S}=\lbrack 1,-1,-1\rbrack$. The 
deterministic reaction rate equation for the concentration $x(t)$ is then governed by the delayed differential 
equation, $\dot{x}={\bf S}\cdot {\bm \nu}=\gamma-\delta\cdot x-\varepsilon\delta \cdot x^{\tau}$. The auxiliary 
coefficient matrices in Eq.~\ref{eq:FokkerPlanck} are the scalars ${\bm \Gamma}=-\delta$ and ${\bf D}=\gamma +\delta 
\cdot x + \varepsilon\delta \cdot x^{\tau}$. For the sake of simplicity, we consider the fluctuations about the 
steady-state $x_s$, where $x_s=\gamma/\delta(1+\varepsilon)$. From Eq.~\ref{eq:FlucDiss}, the variance of the 
fluctuations about $x_s$ is, (using Eq.~\ref{eq:change}),
\begin{gather*}
 \langle{ \left({\frac{X}{V}}\right)^2}\rangle-\langle x_s\rangle^2=\frac{\langle \alpha^2 \rangle}{V} 
=\frac{{\Xi}}{V}=\frac{1}{V}\frac{\gamma}{\delta}.
 \end{gather*}
A useful measure of the relative magnitude of the fluctuations is the fractional deviation $\eta_{\tau}^2$,
\begin{gather}
\eta_{\tau \to \infty}^2=\frac{\Xi}{V\cdot x_s^2}=\frac{1}{N_s}(1+\varepsilon),
\label{eq:fracDev}
\end{gather}
where here $N_s=V\cdot x_s$ is the number of molecules in the steady-state. Without delay, this simple example 
describes a Poisson process, and for a Poisson process the Fano factor $F=\eta^2 N_s=\Xi/x_s=1$. From 
Eq.~\ref{eq:fracDev},
\begin{gather}
F = \left\{ {\begin{array}{*{20}c}
   {1,} & {\tau  \to 0}  \\
   {1 + \varepsilon ,} & {\tau  \to \infty. }  \\
 \end{array} } \right.
 \label{eq:FanoFac}
\end{gather}  
The Fano factor is a particularly convenient statistic to contrast the ordinary and delayed linear noise 
approximations, as well as illustrating the delay time necessary for the present approximation to hold.  
Figure~\ref{fig:Tau} shows the Fano factor $F$ estimated from stochastic simulation~\cite{Gillespie,CaiDSSA} with 
$\gamma=100$, $\delta=4$ and $\varepsilon=0.25$, as compared with the long delay time (solid) and short delay time 
(dotted) estimates. Notice the cross-over occurs for $\tau \approx \delta^{-1}$, that is for delay time comparable to 
the natural time scale of the undelayed kinetics.

The autocorrelation function is given by Eq.~\ref{eq:AutoCorrExpl},
\begin{gather}
K(t)=e^{ - \delta t} \left\{ {1 - \varepsilon \Theta \left( {t - \tau } \right)\delta e^{\delta \tau } \left( {t - 
\tau } \right)} \right\}\frac{{\gamma }}
{{\delta }}.
\end{gather}
This expression coincides with the result of Bratsun {\it et al.}~\cite{Bratsun} to $O(\varepsilon)$, and as they 
demonstrate, $K(t)$ very faithfully reproduces the autocorrelation from simulation data. Furthermore, the 
autocorrelation can be used to identify `quasi-cyles' where regular oscillations emerge from deterministically stable 
systems~\cite{Kuske,Newman,MeTerryBrian,QuasiLimit}.

The delayed-degradation model is used as a transparent illustration of the method, but the same results can be 
obtained by other methods (for example, via moment-generating functions~\cite{Bratsun}). In contrast, the model in 
the next section contains more realistic, nonlinear transition rates, and consequently cannot be treated by existing 
methods. Yet nonlinear rates abound in physical application and exhibit rich dynamics, as the following example 
demonstrates.  

\section{Nonlinear example - Predator-Prey Dynamics}

The methodology outlined in Section~\ref{sec:methods} makes no assumptions about the nonlinearity of the transition 
probabilities in the stochastic model, opening up the possibility to study the dynamics of delayed nonlinear systems. 
As an example capable of exhibiting asymptotic stability and limit cycle behavior, consider the predator-prey model 
with delayed predator birth,
\begin{gather*}
\frac{dP}{dt}=a\cdot P(t)-\frac{a}{K}\cdot P(t)^2-b\cdot P(t)Z(t),\notag\\
\frac{dZ}{dt}=c\cdot P(t-\tau)Z(t-\tau)-d\cdot Z(t),
\end{gather*}  
where $P(t)$ is the density of prey, $Z(t)$ is the density of predators and $K$ is the carrying capacity of the 
environment. Here, $\tau$ is the delay time associated with gestation before the birth of predators. Assuming each 
birth event produces a litter of $1$, the reaction network, in volume $V$, takes the form,
\begin{gather}
\begin{array}{*{20}c}
   P & {\overset{\nu _1 }{\rightarrow}} & {P + 1;} & {} & {\nu _1  = a \cdot \frac{{n_P }}
{V}}  \\
   P & {\overset{\nu _2 }{\rightarrow}} & {P - 1;} & {} &{\nu _2  = \frac{a}
{K} \cdot \frac{{n_P }}
{V} \cdot \frac{{n_P  - 1}}
{V}}  \\
   P & {\overset{\nu _3 }{\rightarrow}} & {P - 1;} & {} & {\nu _3  = b \cdot \frac{{n_P }}
{V} \cdot \frac{{n_Z }}
{V}}  \\
   Z & {\overset{\nu _4 }{\rightarrow}} & {Z - 1;} & {} &{\nu _4  = d \cdot \frac{{n_Z }}
{V}}  \\
   Z & \overset{\nu _5 }{\Rightarrow}  & {Z + 1;} &{} & {\nu _5  = c \cdot \frac{{n_P \left( {t - \tau } \right)}}
{V} \cdot \frac{{n_Z \left( {t - \tau } \right)}}
{V},}  \\
 \end{array}
 \end{gather}
where $n_P/V=P(t)$ and $n_Z/V=Z(t)$. 

With a suitable nondimensionalization,
\begin{gather*}
t_0=1/a,\;\;P_0=K,\;\;Z_0=a/b,\;\;\epsilon=c\cdot K/a,\;\mbox{and}\;\delta=d/a,
\end{gather*}
the deterministic model equations reduce to,
\begin{gather}
\frac{dP}{dt}=P(t)-P(t)^2-P(t)Z(t),\notag\\
\frac{dZ}{dt}=\varepsilon\cdot P(t-\tau)Z(t-\tau)-\delta\cdot Z(t),
\end{gather}
where $\tau$ has been likewise nondimensionalized by $1/a$. The equilibrium point corresponding to coexistence of the 
populations is $(P,Z)=(\frac{\delta}{\varepsilon},1-\frac{\delta}{\varepsilon})$, leading to a necessary condition 
for stable coexistence, with and without delay, that $\delta<\varepsilon$.

It is well-known that delayed rates can have a destabilizing effect on population dynamics~\cite{MayDelay}, and in 
fact can generate limit-cycles in otherwise stable models~\cite{SneppenDelay2,SneppenDelay3,Barrio}. To illustrate 
the approximation method and the destabilizing effects of delay, we consider two values for the delay time, $\tau=0$ 
and $\tau=30$, in two parameter regimes -- the first chosen so that the equilibrium remains asymptotically stable for 
both values of the delay time, the second chosen so that a limit-cycle appears for large delay $\tau$.

\begin{figure}
\includegraphics[width=0.8\linewidth]{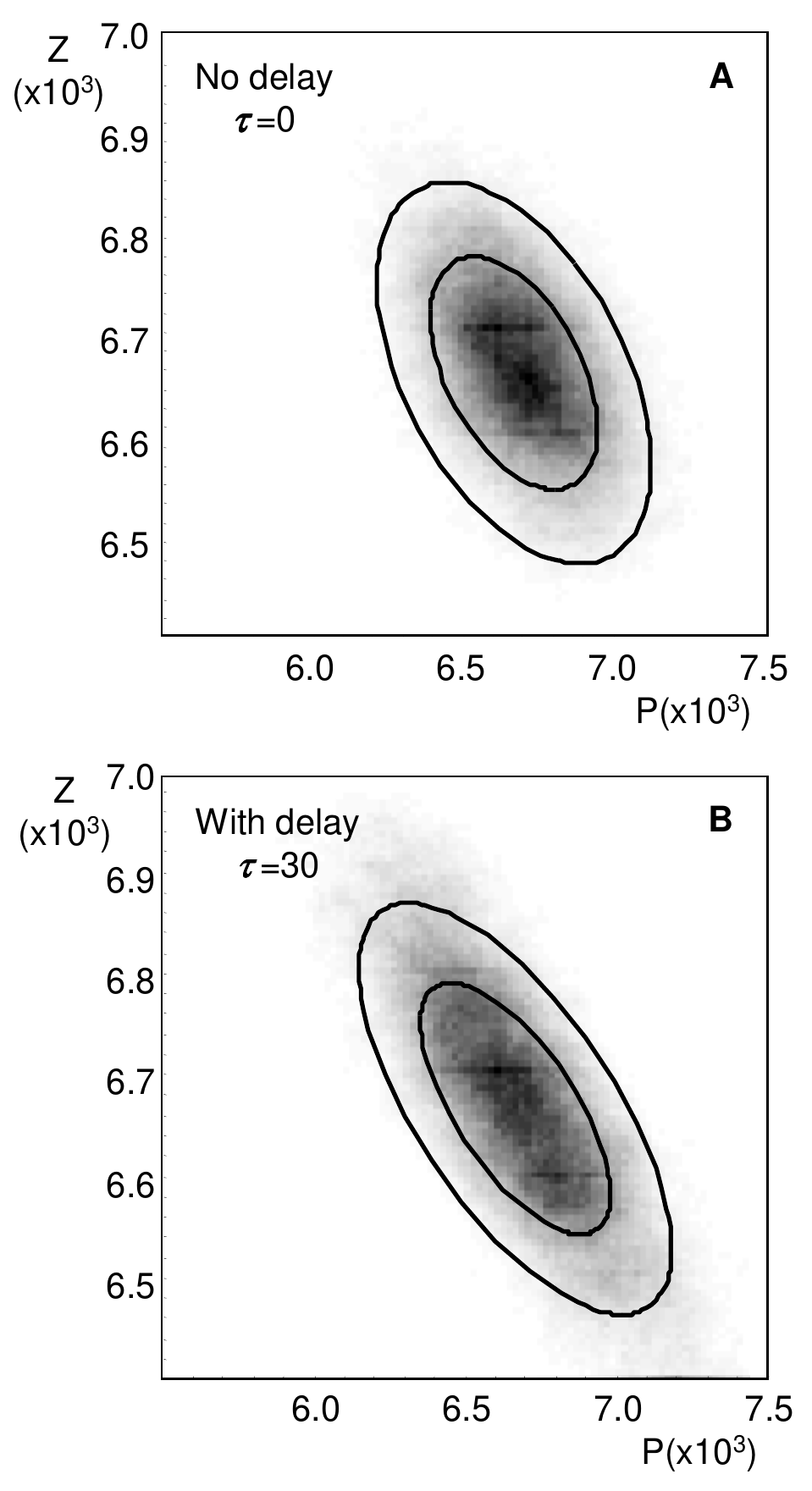}%
\caption{{\bf Steady-state fluctuations in a nonlinear model ($\varepsilon=0.15,\delta=0.05$)}. {\bf A}. Density plot 
of the equilibrium fluctuations from stochastic simulation ($10^6$ realizations). Solid curves correspond to the 
first- and second-standard deviation ellipse computed by an ordinary application of the linear noise 
approximation~\cite{VanKampenME}. {\bf B}. Same model parameters as in panel A, but with delayed predator birth 
($\tau=30$). The solid curves correspond to the first- and second-standard deviation ellipse computed by the delayed 
linear noise approximation.}
\label{fig:NLSteady}
\end{figure}

\subsection{Asymptotically stable}

For ($\varepsilon=0.15,\delta=0.05$), the system remains asymptotically stable in both limits, $\tau=0$ 
(Figure~\ref{fig:NLSteady}A) and $\tau=30$ (Figure~\ref{fig:NLSteady}B). From Eq.~\ref{eq:FlucDiss}, long delay time 
reduces the stability imparted to the system through $\Gamma$. As a consequence, the variance of the fluctuations is 
expected to increase with increasing delay time $\tau$. This increase is evident in the stationary probability 
distribution of the fluctuations derived from stochastic simulation (Figure~\ref{fig:NLSteady}). The ellipses shown 
in the figure correspond to the first- and second-standard deviations of the steady-state Gaussian distribution 
predicted by the approximation, while the density plot represents extensive stochastic simulation data generated 
using Gillespie's algorithm~\cite{Gillespie,CaiDSSA}.

The most striking consequence of the delay on the intrinsic fluctuations is the increased magnitude of the 
cross-correlation between $P$ and $Z$. As $\tau\to \infty$, the delayed rate $c\cdot P(t-\tau)Z(t-\tau)$ no longer 
offers compensation to the predation event with rate $-b\cdot P(t)Z(t)$, resulting in the increased 
cross-correlation. This is an example of the nontrivial effect of delayed dynamics on intrinsic fluctuations, even 
though the equilibrium point is stable. In situations where the delay affects not only the fluctuations, but the 
underlying stability itself (as is the case for delay induced limit cycles), the analysis becomes more complicated.

\begin{figure}
\includegraphics[width=0.8\linewidth]{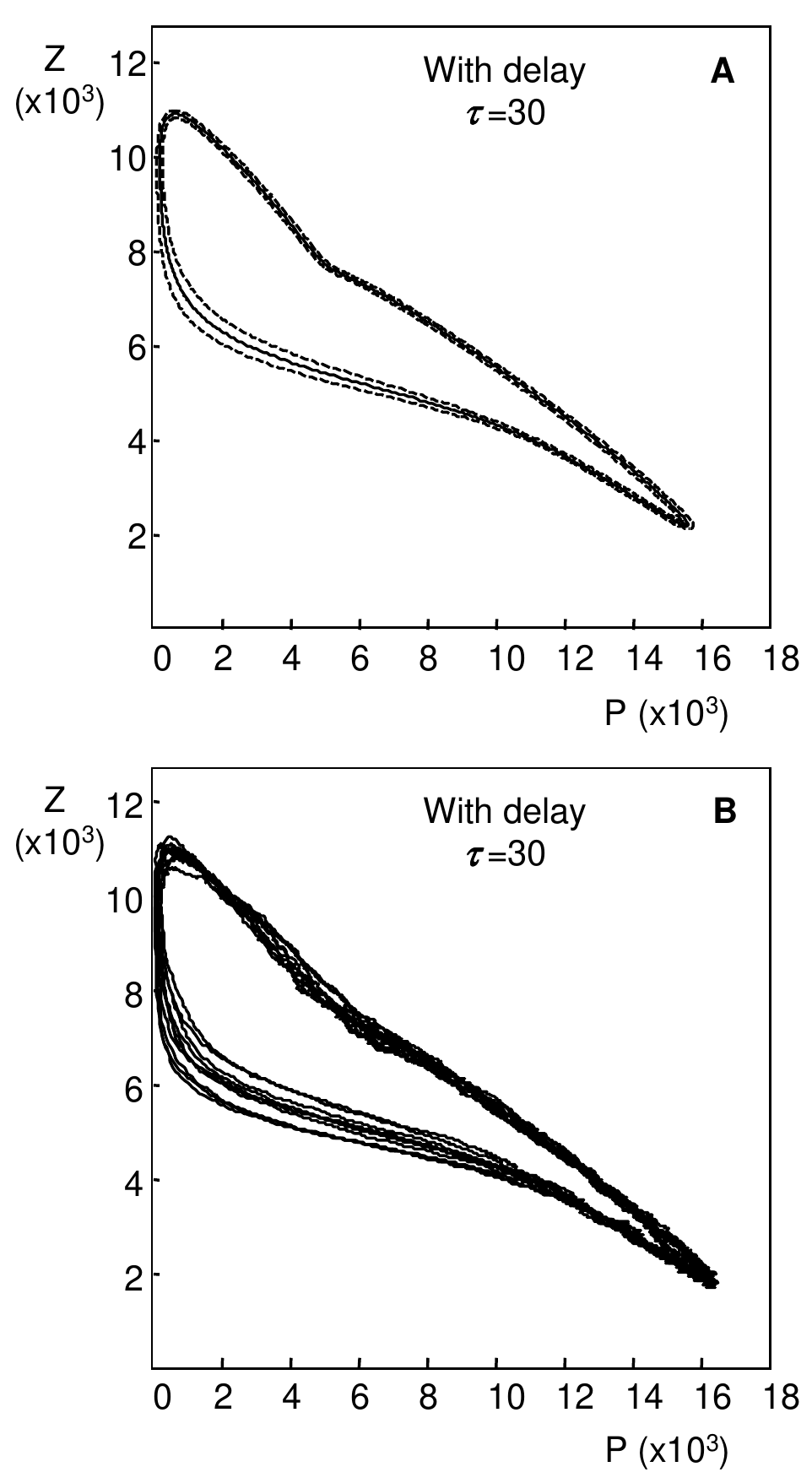}%
\caption{{\bf Fluctuations around a delay-induced limit cycle ($\varepsilon=0.25,\delta=0.05$)}. {\bf A}. Envelope of 
the standard-deviation of the fluctuations transverse to the limit cycle, as computed by Eq.\ref{eq:LimCorr} using 
the delayed linear noise approximation. {\bf B}. Stochastic simulation of the system along the limit cycle.}
\label{fig:NLLimit}
\end{figure}

\subsection{Limit cycle}

For ($\varepsilon=0.25,\delta=0.05$), the system is asymptotically stable for $\tau=0$, but a limit-cycle appears for 
$\tau=30$. By separating the fluctuations tangent to the limit cycle from those 
transverse~\cite{LimitCycle2,LimitCycle1,MeBrianMads,QuasiLimit}, the delayed linear noise approximation is easily 
extended to a system exhibiting a limit-cycle.

Briefly, a moving coordinate frame is introduced using as a basis the unit vectors tangent (${\bm \hat s}$) and 
normal (${\bm \hat r}$) to the limit cycle. In the moving frame, the covariance of the transverse fluctuations 
$\Xi_{rr}$ decouples from the divergent fluctuations along the limit cycle, and is characterized by a stable 
evolution equation ({\it cf.} Eq.~\ref{eq:LNACorrEv}),
\begin{gather}
\frac{d\Xi_{rr}}{dt}=2\Gamma'_{rr}\;\Xi_{rr}+D'_{rr},
\label{eq:LimCorr}
\end{gather}
where $\Gamma'_{rr}$ and $D'_{rr}$ are elements of the drift and diffusion matrices in the moving frame,
\begin{gather}
{\bf \Gamma}'={\bf U}\cdot {\bf \Gamma} \cdot{\bf U}^{\dagger}+\frac{d{\bf U}}{dt}\cdot {\bf U}^\dagger,\;\;\;{\bf 
D}'={\bf U}\cdot {\bf D}\cdot {\bf U}^\dagger,
\end{gather}
and ${\bf U}$ is the rotation matrix generated from the deterministic rate equations ${\bf f}({\bf x},{\bf x}^\tau)$,
\begin{gather}
{\bf U}=\frac{1}{\sqrt{f_1^2+f_2^2}}\left[ {\begin{array}{*{20}c}
   {f_1 } & { - f_2 }  \\
   {f_2 } & {f_1 }  \\
 \end{array} } \right].
\end{gather}

Figure~\ref{fig:NLLimit} illustrates the estimate of the fluctuations along the delay-induced limit cycle via the 
delayed linear noise approximation (Figure~\ref{fig:NLLimit}A), compared with the result of a stochastic simulation 
(Figure~\ref{fig:NLLimit}B). The width of the envelope of the fluctuations is not uniform around the orbit, 
reflecting the state-dependent drift ${\Gamma}$ and diffusion ${\bf D}$ matrices in Eq.~\ref{eq:LNACorrEv}. This same 
non-uniformity is also observed in the stochastic trajectory. 

The nonlinear predator-prey model demonstrates the utility and comparative simplicity of the delayed linear noise 
approximation -- once the network is written in terms of the stoichiometry matrix and the propensity vector, despite 
the lengthy derivation, Eqs.~\ref{eq:delayFP}, \ref{eq:LNACorrEv} and \ref{eq:AutoCorrExpl} allow algorithmic 
characterization of the fluctuations.

\section{Discussion}
In models of cellular chemical reaction systems, spatial transport and long auxiliary pathways are often represented 
using time-delayed reaction rates. At the mesoscopic level, delayed dynamics result in a probability conservation 
equation that characterizes a non-Markovian process. Since analytic solutions are rare, approximation of the 
governing equations are necessary. In the limit of large numbers of molecules, weak delayed feedback and long delay 
time, we have derived the leading order behavior of a probability conservation equation with delayed transition rates 
from an expansion in the system volume $V$. The fluctuations are characterized by a linear Fokker-Planck equation, in 
accordance with the linear noise approximation of the undelayed case~\cite{VanKampenME}, and coinciding with a 
delayed random walk in a quadratic potential~\cite{MiltonBOOK}. We find that the delayed dynamics contribute unevenly 
to the drift and diffusion coefficients of the Fokker-Planck equation, and conclude that long time-delay can only 
increase the magnitude of intrinsic fluctuations for systems where the delayed feedback has a stabilizing effect.

Here, we have focused upon two example systems -- one that evolves toward a stable steady-state, the second is a 
nonlinear model exhibiting a delay-induced limit cycle. It is often the case that models with delayed rates are used 
to describe oscillatory dynamics~\cite{SneppenHes1}. The delayed linear noise approximation is easily adapted to 
systems evolving along a stable limit cycle by a simple change of coordinates~\cite{LimitCycle1,MeBrianMads}. 

Finally, the effect of noise on the macroscopic behavior of a system is not always additive, and in fact noise can 
generate ordered oscillations from a deterministically stable model~\cite{Bratsun,Constructive}. These noise induced 
oscillations have been proposed as a mechanism to extend the capacity of a given network to sustain 
oscillations~\cite{Elowitz2,Vilar}. The results derived above, specifically the autocorrelation function 
Eq.~\ref{eq:AutoCorrExpl}, allow the method developed for studying noise-induced oscillations in undelayed systems to 
be applied to systems characterized by delayed dynamics~\cite{MeTerryBrian}.

Funding was provided by Canada's NSERC Post-doctoral fellowship, support through NSF Grant No. MCB0417721, and by 
Grant Nos. PHY-0216576 and PHY-0225630 through the PFC-sponsored Center for Theoretical Biological Physics during a 
post-doctoral stay. The author thanks Jian Liu, Stefan Klumpp, Peter McHale, Terry Hwa, Sue Ann Campbell and Lev 
Tsimring for many helpful comments.

\bibliographystyle{apsrev}

\end{document}